\documentclass{ws-ijmpa}

\begin{document}

%
%

\title{$\pi\pi$ AMPLITUDES FITTED TO EXPERIMENTAL DATA AND TO ROY'S
EQUATIONS}

\author{\footnotesize R. KAMI\'NSKI, L. LE\'SNIAK}

\address{Henryk Niewodnicza\'nski Institute of Nuclear Physics, Polish Academy of
Sciences,\\ 
 ul. Radzikowskiego 152, PL 31-342, Krak\'ow, Poland}

\author{B. LOISEAU}

\address{LPNHE\thanks{Unit\'e de Recherche des Universit\'es Paris 6 et Paris 7, associ\'ee au CNRS.}, 
       Universit\'e P. et M. Curie, 4, Place Jussieu, \\
         75252 Paris Cedex 05, France}

\maketitle


\begin{abstract}
The scalar-isoscalar, scalar-isotensor and vector-isovector $\pi\pi$ amplitudes 
are fitted simultaneously to experimental data and to Roy's equations.
The resulting amplitudes are compared with those fitted only to experimental data.
No additional constraints for the $\pi\pi$ threshold behaviour of 
the amplitudes are imposed.
Threshold parameters are calculated for the amplitudes in the three waves. 
Spectrum of scalar mesons below 1.8~GeV is found 
from the analysis of the  analytical structure of the fitted amplitudes.

\end{abstract}

\vspace{0.3cm}

The $\pi\pi$ amplitudes in the scalar-isoscalar ($S0$), the scalar-isotensor ($S2$) 
and the vector-isovector ($P1$)
waves have already been analyzed by us with the help of Roy's equations~\cite{kll03}.
Our aim was to eliminate the "up-down" ambiguity in the $S0$ phase shifts below 1~GeV~\cite{klr97}.
We have shown that only amplitudes reproducing the
"down-flat" data set fulfill Roy's equations.
Therefore they satisfy crossing symmetry and should then correspond to physical amplitudes.

The aim of this talk is to present preliminary results in a construction of the theoretical $\pi\pi$ amplitudes which 
simultaneously describe the experimental data and fulfill Roy's equations up to 1150
MeV.
In order to study the constraints from Roy's equations we also calculate amplitudes 
fitted only to experimental data in the three channels ({\it fit 1}).
The analytical form of the $S0$ wave has been obtained in the coupled three 
channel model~\cite{kll99}.
Now a new fit to the "down-flat" data from the $\pi\pi$ threshold up to 1600~MeV is performed. 
In the $S2$ wave we have used the same model as in~\cite{kll03} and the data corresponding to the solution A of Hoogland et al.~\cite{hoogland77}. 
In the $P1$ wave we have introduced 
 a  two channel model with  a  rank-two and a rank-one separable potentials in 
the $\pi\pi$ and $K\bar{K}$ channels, respectively.
Amplitude of the $P1$ wave has been fitted to the data of Hyams~{\it et al.}~\cite{Hyams73} in the
effective two pion mass $m_{\pi\pi}$ from 600 up to 1400~MeV.
 The numbers of free parameters in the $S0$, $S2$ and $P1$ waves are 14, 4 and 8, respectively.

The total $\chi^2$ used in the simultaneous fit to experimental data and to Roy's equations ({\it fit 2})
has been defined in our previous work~\cite{kll03}. It consists of two parts: $\chi^2_{exp}$ - 
related to the experimental data and $\chi^2_{Roy}$ - related to Roy's equations. 
The latter one is defined as the sum of three terms for the isospin $I=0,1,2$ in the $s=m_{\pi\pi}^2/m_{\pi}^2$ 
range from 4.1 up to 67.
The 23 points in the calculations of $\chi^2_{Roy}$ have been  chosen as follows: $s_1=4.1, \,\,\, s_2 = 5$ and $s_j = 3j-2$ for 
$j=3,4 ...23$.

In Table~1 the $\chi^2$ components for the experimental data and for Roy's equations are presented.
It can be seen that while the data for the phase shifts $\delta^{I=0}_{J=0}$ and inelasticity ${\eta^0_0}$ in 
the $S0$ wave are similarly well described in both fits, 
the  data for  $S2$, $P1$ and the data~\cite{cohen80} for the sum ${\delta_{\pi\pi}+\delta_{K\bar{K}}}$  are 
 better reproduced in {\it fit~1}. 
One can conclude that some parts of these data 
 do not satisfy the constraints imposed by crossing symmetry.

The real parts of the amplitudes from {\it fit~1} and {\it fit~2} are shown in Fig.~1. 
Again one can see that  the  agreement of the experimental data in the $S2$ and $P1$ waves with the  
amplitudes of the {\it fit~2} is  not as good as that of the {\it fit~1}. 
In the {\it fit~2} the averaged values of differences between real parts of amplitudes used as input in Roy's equations and real 
parts of amplitudes obtained as output are  smaller than (2, 1 and $ 3)\times 10^{-3}$ for the 
$S0$, $S2$ and $P1$ waves, respectively.

\vspace{-0.3cm}

\begin{table}[h!]
\tbl{Values of various components of $\chi^2_{tot}$ with the corresponding numbers of data points in
parentheses. References to experimental data can be found in$\,^1$. Symbol $\phi$ denotes sum 
${\delta_{\pi\pi}+\delta_{K\bar{K}}}$.}
{\begin{tabular}{@{}ccccccccccccc@{}} 
Fit & $\chi^2_{tot}$ & $\chi^2_{Roy^0_0}$ & $\chi^2_{Roy^2_0}$ & $\chi^2_{Roy^1_1}$ & $\chi^2_{Roy}$ &
$\chi^2_{\delta^0_0}$ & $\chi^2_{\eta^0_0}$ & $\chi^2_{\phi}$ &
$\chi^2_{\delta^2_0}$ & $\chi^2_{\delta^1_1}$ & $\chi^2_{\eta^1_1}$ & $\chi^2_{exp}$ \\
  & (248) & (23)   & (23)   & (23)   & (66) & (56)  & (30) & (21) & (12) & (40) & (20) & (179)\\
1 & 123.5 & ----- & ----- & ----- & ----- & 48.7 & 28.7 & 10.7 &  1.9 & 20.1 & 13.4 & 123.5\\
2 & 247.5 &  7.1 &  2.7 & 16.0 & 25.8 & 59.7 & 30.3 & 33.9 & 26.2 & 49.0 & 22.6 & 221.7\\
\end{tabular}}
\end{table}

\vspace{-0.4cm}

In Table~2 values of the scattering lengths $a^I_J$ and the slope parameters $b^I_J$ as defined in our paper~\cite{kll03} 
and positions of the most important $S$-matrix poles~\cite{kll99} for the $S0$ wave 
are presented. 
In the {\it fit 1} the threshold parameters of the  $S2$ and $P1$ amplitudes are much different from those 
calculated for the {\it fit 2}. 
In Fig.~1 one can see significant differences in the $\pi\pi$ threshold behaviour of the  $S2$ and $P1$ 
amplitudes.
This is due to a lack of experimental data near the $\pi\pi$ threshold in these waves 
and a lack of constraints from Roy's equations in the {\it fit~1}.
Note also the differences in the positions of the poles. 
Positions of the  $\rho$ meson poles depend also on the fit. 
They have been calculated for $Im\,k_{\pi}<0,\,\,\,Im\,k_{K}>0$ and are $(764 - i55)$ MeV and $(765 - i72)$~MeV 
for the {\it fit~1} and {\it fit~2}, respectively.

\vspace{-0.3cm}

\begin{table}[h!]
\tbl{Values of $a^I_J$, $b^I_J$ and pole positions for the $S0$ wave in the complex energy plane (in~MeV) on sheets denoted by 
signs of imaginary parts of the momenta in three channels}
{\begin{tabular}{@{}cccccccccc@{}} 
Fit & $a^0_0$ & $b^0_0$ & $a^2_0$ & $b^2_0$ & $a^1_1$ & $b^1_1$ & $(-,+,+)$ & $(-,+,+)$ & $(-,-,-)$  \\
1 & 0.210 & 0.255 & -0.097 & -0.064 & 0.32  & -0.017 & $475-i242$ & $1008-i32$ & $1415-i85$  \\
2 & 0.204 & 0.243 & -0.044 & -0.076 & 0.047 & 0.0072 & $450-i280$ &  $981-i50$ & $1474-i73$  \\
\end{tabular}}
\end{table}

\vspace{-0.4cm}

Work to consider, in our $\pi \pi$ amplitudes, QCD low energy
constraints from Chiral Perturbation Theory is in progress.

\newpage

\begin{figure}[t]
\centerline{\psfig{file=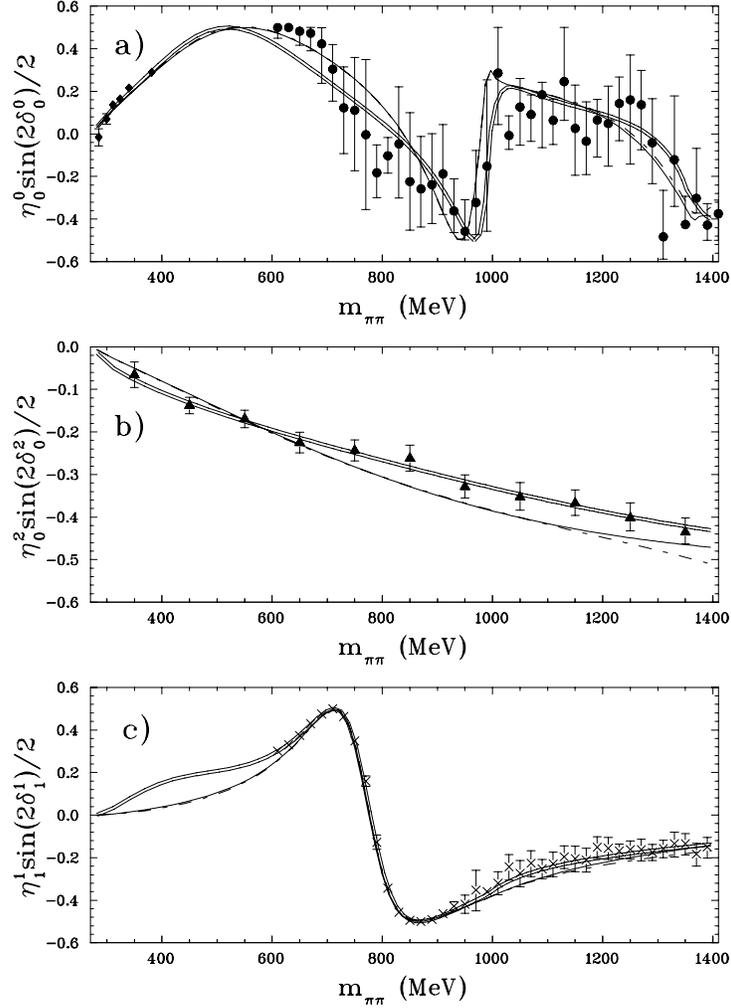,width=9.5cm}}
\vspace*{8pt}
\caption{Real parts of the $\pi\pi$ amplitudes multiplied by $2k/m_{\pi\pi}$ 
($k$ is the relative momentum in the $\pi\pi$ system).
Double solid lines correspond to {\it fit~1}.
Single solid and dashed lines correspond to the "input" and the "output"  amplitudes, respectively for the
{\it fit~2}. 
References to experimental data can be found in$\,^1$.}
\end{figure}

\end{document}